\documentclass{amsart}
\newcommand\Cech{\v{C}ech\xspace}
\usepackage{graphicx,amssymb,amsmath}
\usepackage[ruled]{algorithm2e}
\usepackage{caption}
\usepackage{subcaption}

\usepackage{hyperref}

\usepackage[style=numeric,backend=bibtex]{biblatex}
\usepackage{amsaddr}

\begin{document}

\title{An improved algorithm for Generalized \v{C}ech complex construction}

\author{Jie Chu$^{1}$, Mikael Vejdemo-Johansson$^{1,2}$ and Ping Ji$^{1,3}$}
\email{jchu1@gradcenter.cuny.edu, mvj@math.csi.cuny.edu, pji@gc.cuny.edu}

\address{$^1$ CUNY Graduate Center, 365 Fifth Avenue, New York, NY, USA}
\address{$^2$ CUNY John Jay College, 524 W 59th St, New York, NY, USA}
\address{$^3$ CUNY College of Staten Island, 2800 Victory Blvd, Staten Island, NY, USA}

\begin{abstract}
	In this paper, we present an algorithm that computes the generalized \Cech complex for a finite set of disks where each may have a different radius in 2D space. An extension of this algorithm is also proposed for a set of balls in 3D space with different radius. 
	To compute a $k$-simplex, we leverage the computation performed in the round of $(k-1)$-simplices such that we can reduce the number of potential candidates to verify to improve the efficiency. An efficient verification method is proposed to confirm if a $k$-simplex can be constructed on the basis of the $(k-1)$-simplices. We demonstrate the performance with a comparison to some closely related algorithms.
\end{abstract}


\maketitle

\section{Introduction}
\label{sec:intro}
In topological data analysis, several types of simplicial complexes that encode geometric information are in widespread use. Although for computational reasons the Vietoris-Rips complex is most commonly used, the \Cech complex has much stronger theoretical justifications. For instance, \cite{niyogi2008finding} proved that a \Cech complex can recover the homological structure of a manifold given dense enough sampling from the manifold and \cite{de2007coverage,de2007homological} proved that the sensor network coverage problem can be solved using the connectivity information in a \Cech complex.

In this work we are interested in the generalized \Cech complex: we are given a set of disks $D$ in the 2D plane, where each disk $d_i$ has its own radius $r_i$. The vertices of the \Cech complex are formed by the disk centers, and a $k$-simplex is included in the generalized \Cech complex precisely when all $k+1$ disks associated with the vertices share a common point of intersection. This is a stricter criterion than for the Vietoris-Rips complex, where it is sufficient that each pair of disks has a non-empty intersection. The Alpha complex is also homotopy equivalent to the union of disks. It is defined over the set of cells which are the intersection of disks and its Voronoi cells\cite{edelsbrunner2010alpha}. 

\Cech complex is excellent in capturing the exact topology for applications like the network coverage modeling, where blanket coverage problem can be addressed easily by the \Cech complex\cite{de2007coverage}. It also reveals higher order information about the coverage, such as the level of coverage or redundant sensors in the network. Even though the Alpha complex is homotopy equivalent, the boundary groups carry information about redundancy in the network that requires the \Cech complex to fully measure it.

But the stricter definition also leads to great difficulty in computation. There has been some research in computing the \Cech complex. If all disks have the same radius, which is called the standard \Cech complex, \cite{dantchev2012efficient} proposes an efficient algorithm, and many software packages are available\cite{dionysus,gudhi,cechmate}. \cite{le2015construction} outlines an algorithm for computing the generalized \Cech complex for a set of disks with varying radius and \cite{espinoza2020numerical} proposes a method to compute the \Cech complex filtration over a set of disks. 

Our work is inspired by the excellent work in~\cite{le2015construction,espinoza2020numerical}, and we propose an improved algorithm based on the work in \cite{le2015construction}. In section~\ref{sec:algo} we describe the algorithm for computing generalized \Cech complex in both 2D and 3D. In section~\ref{sec:exp}, we perform experiments for evaluating the performance of our algorithm in comparison to the work in \cite{le2015construction}. We summarize and conclude the paper with possible future works in section~\ref{sec:conclusion}. 

\section{Background and Related work}
\label{sec:background}
In $\mathbb{R}^d$, \Cech complex $C$ is defined on a set of points $P$, with $P$ as the vertex set of the \Cech complex. A $k$-simplex is in $C$ if the $d$-balls centered at $k+1$ points have a non-empty intersection.
Given a set of disks $D$ in $\mathbb{R}^2$, where each disk $d_i \in D$ has its center located at $p_i = (x_i, y_i)$, with radius $r_i$, there are several common choices of simplicial complex to encode the geometric information of the set of disks. When $r_i = \frac{\epsilon}{2}, \forall i$, we can define the standard \Cech complex as $C_\epsilon(D) = \{\sigma \subseteq D | \bigcap_{d \in \sigma} d \neq \emptyset\}$. The generalized \Cech complex is defined as the same, but without the condition that all $r_i$ have the same value, so that each circle may have different radius. This definition can be easily extended to 3D space, as we will see in Section~\ref{subsec:algo3d}, we will focus on 2D space in the discussion of this section. 

Due to the difficulty of computation and the size of its representation, \Cech complex is often approximated with other simplicial complexes or techniques \cite{kerber2013approximate}.
The Vietoris-Rips complex is a common choice to approximate the \Cech complex as in \cite{de2007coverage}, due to its lower computation cost. The Vietoris-Rips complex can be defined as $\textit{VR}(D) = \{\sigma \subseteq D | \|d_i - d_j\| < r_i+r_j, \forall d_i, d_j \in \sigma \}$, where $\|d_i - d_j\| $ denotes the Euclidean distance of the centers of the two disks, i.e. a set of disks form a $k$-simplex if they form a $k+1$-clique in the neighborhood graph. 

The \Cech complex and the Vietoris-Rips complex agree on the 0 and 1 simplices, and the \Cech complex is a subcomplex of the Vietoris-Rips complex when $k \geq 2$. \cite{zomorodian2010fast} formulated a two-phase approach to compute Vietoris-Rips complexes efficiently. The first phase is to construct the neighborhood graph which is the 1-skeleton of both \Cech complex and Vietoris-Rips complex. \cite{zomorodian2010fast} surveyed several methods for this process and compared their performances. In this paper, we do not study the construction of the neighborhood graph further.

The second phase in their work is to construct the Vietoris-Rips complex incrementally. In this paper, we employ a similar approach as what Zomorodian described as the ``inductive algorithm'' to enumerate candidates for a potential \Cech $k$-simplex to be verified. However, the \Cech complex has a stricter definition compared to the Vietoris-Rips complex, so that a Vietoris-Rips $k$-simplex is merely a candidate for a \Cech $k$-simplex, requiring further verification as proposed in Section~\ref{sec:algo}. 
In verifying a candidate for the \Cech complex, a closely related algorithm as used in \cite{dantchev2012efficient} is the minimum enclosing ball(/sphere) algorithm. In the standard \Cech complex setting, to determine if there is a non-empty intersection among multiple disks, it can be formulated as whether the radius of the minimum enclosing ball of the centers of disks is less than the radius of disks. There are many studies on the minimum enclosing ball problem and an efficient algorithm can be found in \cite{gartner1999fast}.

The Alpha complex($\alpha$-complex) is also homotopy equivalent to the union of disks and widely used\cite{edelsbrunner2010alpha}. For each disk $d_i$ centered at $p_i$, the Voronoi cell is defined as $V_i = \{ x \in R^2 | \|x-p_i \| \leq \|x-p_j\|, \forall j\}$. 
We can define $R_p(d_i) = d_i \cap V_i$, and the alpha complex is defined as $Alpha(D) = \{\sigma \subseteq D | \bigcap_{d_i \in \sigma} R_p(d_i) \neq \emptyset\}$. A weighted version can be defined by assigning weights to each disk, similar to the generalized \Cech complex. \cite{sheehy2015output} proposes an output-sensitive algorithm to compute weighted alpha-complex.
However, as we argued in the introduction, \Cech complex retains the geometric information about the redundancy of coverage better when we use it to model sensor networks. The boundary groups are much smaller when using alpha complexes than when using \Cech complexes. Hence, the alpha complex will give a misleading notion of how redundant the network really is. 

Given a set of curves, there are algorithms\cite{toth2017handbook} that compute 2D arrangements, which decompose the plane into open cells, and thus can be used for constructing \Cech complex. An open cell bounded by multiple curves is called a face. One can compute the 2D arrangement given the set of the circular boundaries of all disks. For a face $F$, any subset of $D_F$, where $D_F = \{d \in D | d \supset F\}$, is corresponding to a \Cech simplex. However, there may be some faces which are not corresponding to the intersection region of a \Cech simplex and a simplex may be related to multiple faces. Due to this over-counting issue plus the time complexity of the 2D arrangement algorithm itself, the efficiency of this approach to construct the \Cech complex is yet to be evaluated.

\section{Algorithms}
\label{sec:algo}
We will first deal with 2D space. In $\mathbb{R}^2$, we are given a set of disks $D$, where disk $d_i \in D$ has its center located at $(x_i, y_i)$, with radius $r_i$, we want to construct the generalized \Cech complex for $D$ as defined in \ref{sec:background}. The core task is to determine if a set of disks has a non-empty intersection, i.e., for a set of disks $\{d_0, d_1, ..., d_k\}$, if $\bigcap_{i=0}^{k} d_i \neq \emptyset$ then they form a $k$-simplex. Both 0-simplices and 1-simplices are easy to compute, with a 0-simplex as the center of each disk and the 1-simplices corresponding to the edges of the neighborhood graph. We will focus on the case when $k>2$.

One key observation of the \Cech complex construction process is that we could obtain a $k$-simplex by extending a $(k-1)$-simplex. A $k$-simplex should contain $k$ disks that form $(k-1)$-simplex, and a disk that at least has non-empty intersections with all these existing $k$ disks. In this way, we can iteratively compute it based on the previous calculation.

\begin{figure*}[htbp]
	\centering
	\begin{subfigure}[b]{0.3\textwidth}
		\centering
		\includegraphics[width=\textwidth]{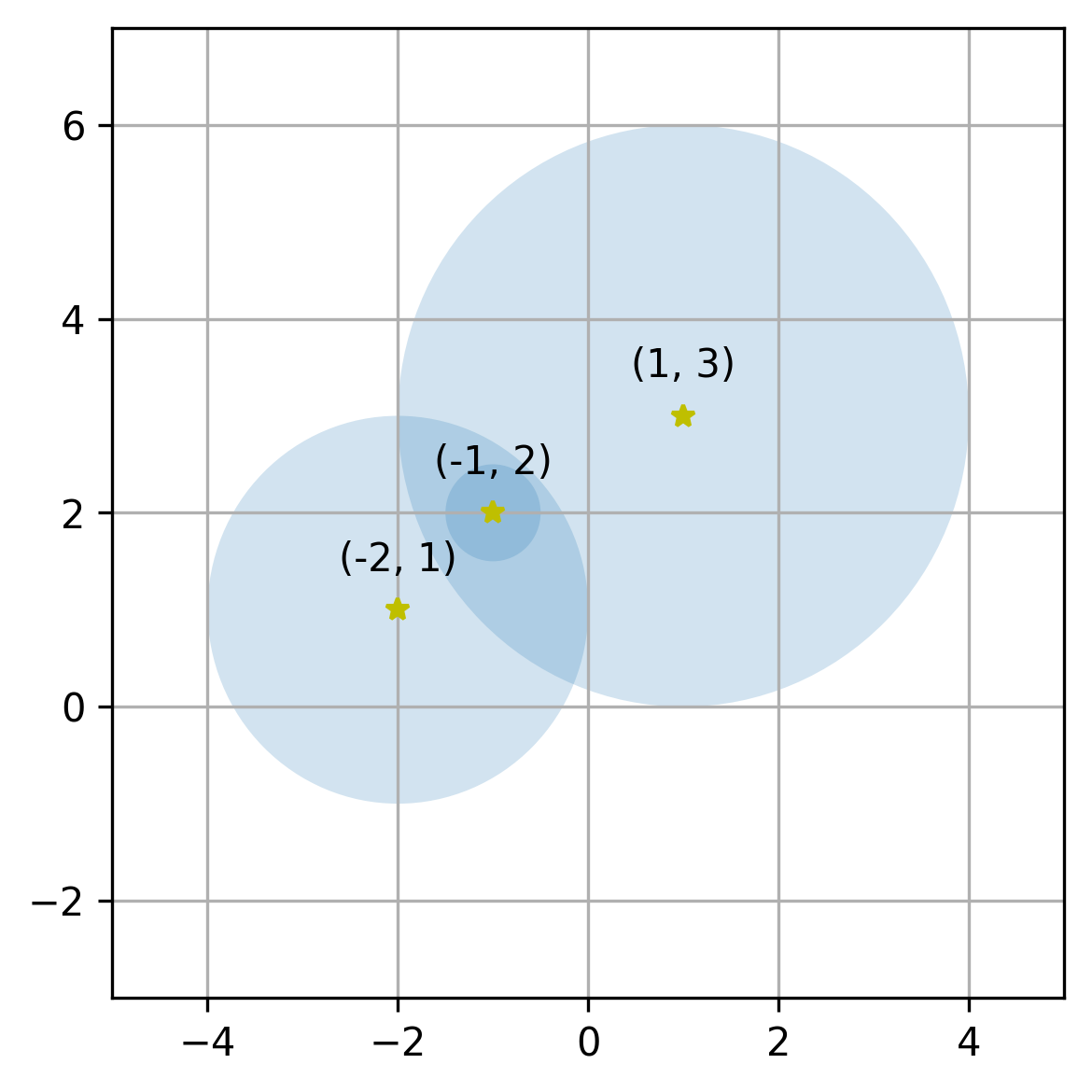}
		\caption{The smallest disk which is inside all other two disks}
	\end{subfigure}
	\hfill
	\begin{subfigure}[b]{0.3\textwidth}
		\centering
		\includegraphics[width=\textwidth]{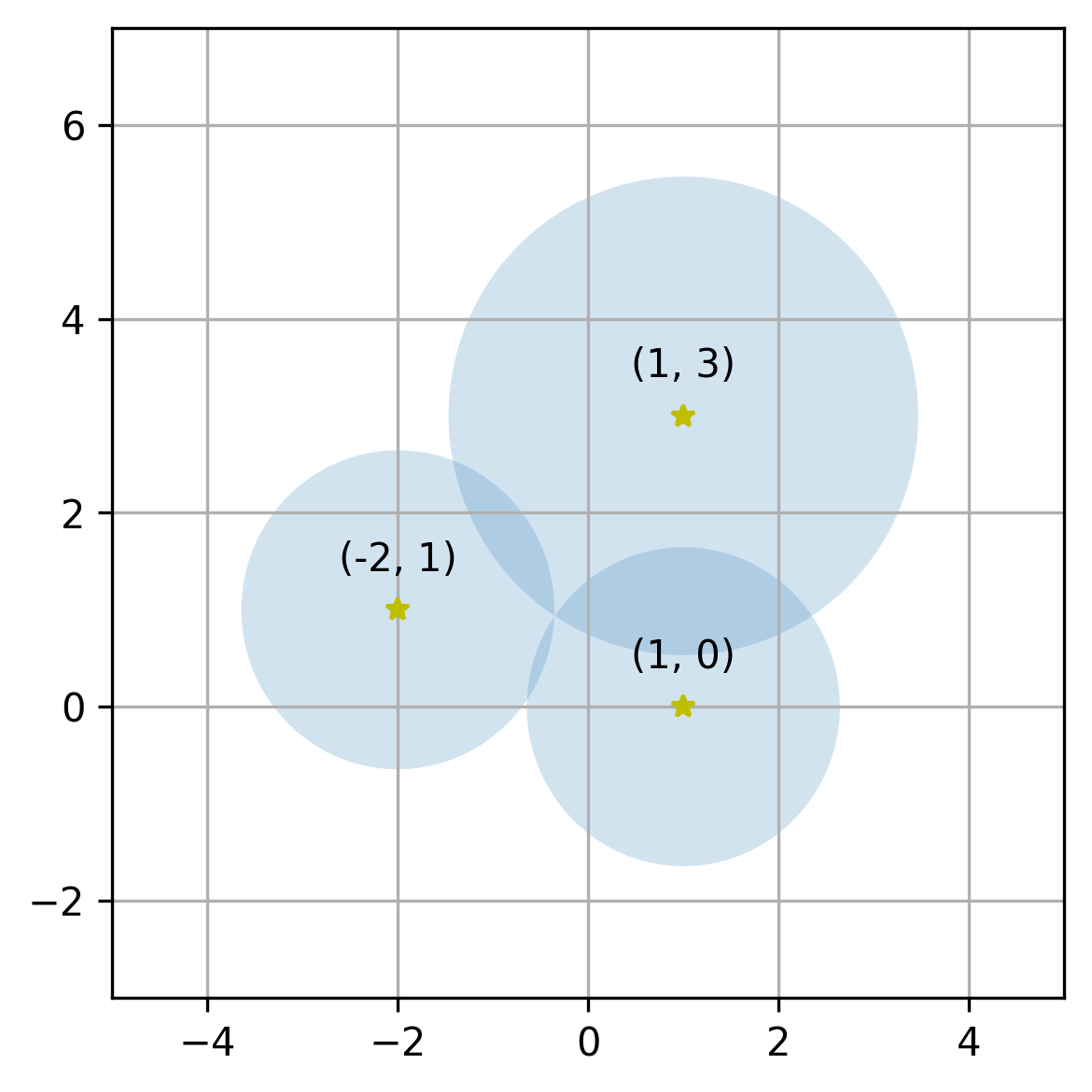}
		\caption{Intersect at a single point}
	\end{subfigure}
	\hfill
	\begin{subfigure}[b]{0.3\textwidth}
		\centering
		\includegraphics[width=\textwidth]{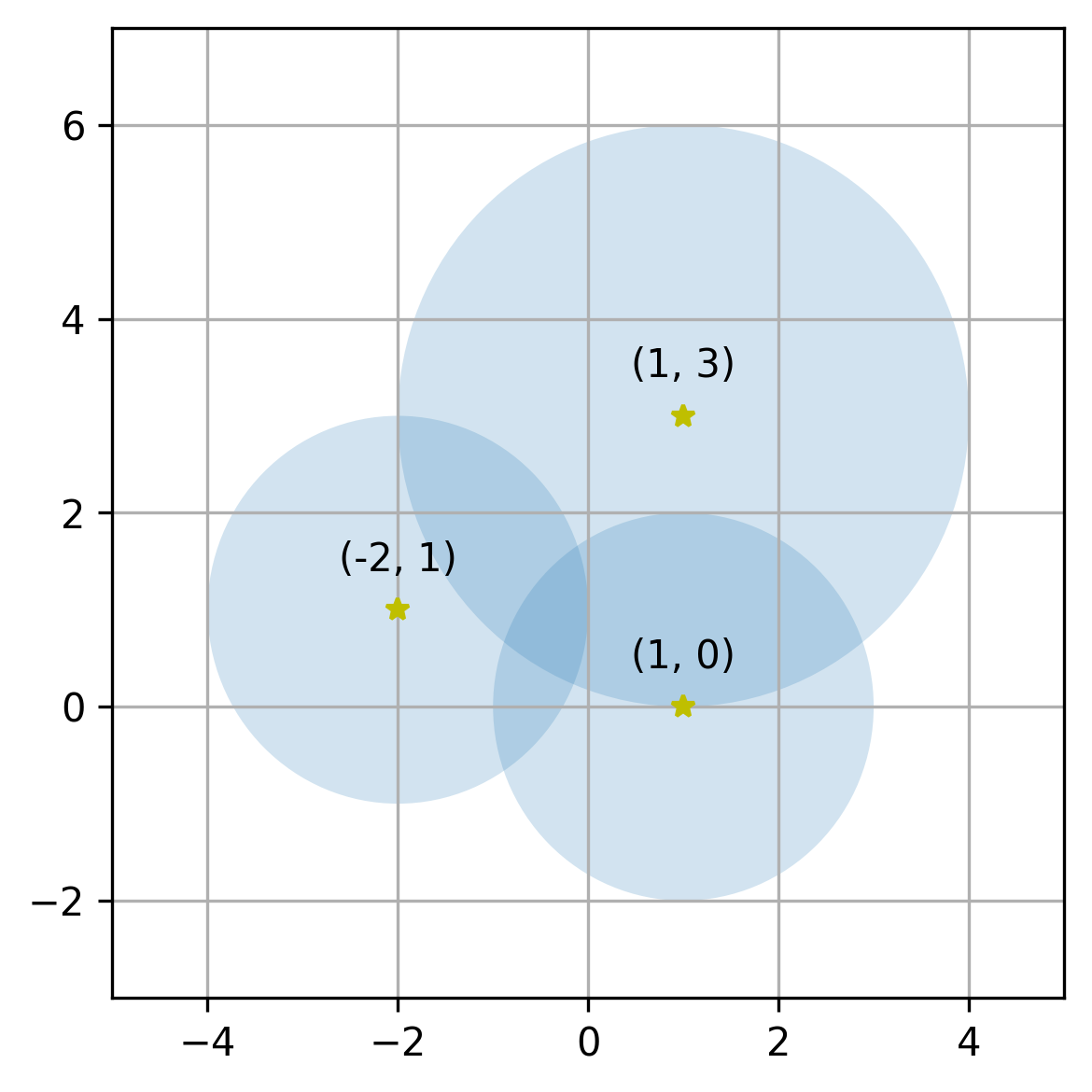}
		\caption{Intersect with a region bounded by multiple circumference arcs}
	\end{subfigure}
	\hfill
   \caption{Intersections of circles}
   \label{fig:intersection}
\end{figure*}

Once we are given a set of $k+1$ disks as a potential candidate for $k$-simplex, we need to verify all disks have a non-empty intersection. As discussed in \cite{espinoza2020numerical}, the intersection for a $k$-simplex is one of the following scenarios (as shown in Figure~\ref{fig:intersection} for three circles):
\begin{enumerate}
	\item The smallest disk which is inside all other $k$ disks,
	\item A single point,
	\item A region bounded by multiple circumference arcs.
\end{enumerate}

The first scenario can be verified efficiently. 
The second scenario is a special case of the third. As of the third scenario, the intersection region is bounded by multiple circumference arcs, that each one is part of the boundary of a disk. We can define this set of disks as the boundary support set $B(s_k)$ for $k$-simplex $s_k$. The boundary of two disks will have two intersection points. We will justify in the following part that it is possible to determine if the intersection is non-empty by verifying intersection points of some disk boundaries.
 
As described earlier, we are given a $(k-1)$-simplex $s_{k-1} = \{d_{i_0}, d_{i_1}, ..., d_{i_{k-1}}\}$ and a new disk $d_{i_k}$ to be determined if they form a $k$-simplex. If they do form a $k$-simplex $s_k$, either the intersection of the disks in the $(k-1)$-simplex is completely inside the newly added disk $d_{i_k}$, such that the support set doesn't change, or $d_{i_k}$ reduces the intersection region such that it belongs to the support set $B(s_k)$ and part of its boundary bounds intersection for the $k$-simplex. And thus we can simply verify if either of those two cases is true. 

For the first case, we can verify if the intersection point that we computed for $(k-1)$-simplex lies in the newly added disk $d_{i_k}$. And for the second case, since part of the boundary of the newly added disk $d_{i_k}$ is in the set of the arcs that bounds intersection for the $k$-simplex, then there has to be an intersection point $p$ between the boundary of $d_{i_k}$ and the boundary of some other disk $d_{i_j} \in B(s_k)$ that lies in the range of all disks. We'll denote intersection points of the boundaries of two disks $d_i$ and $d_j$ as $p\in\partial d_{i}\cap\partial d_{j}$. Since we don't know $B(s_k)$, we will simply verify for every disk in $s_{k-1}$, if any of their intersection points with the newly added disk $d_{i_k}$ is inside all other disks.
\begin{algorithm}[htbp]
	\SetAlgoLined
	\SetKwProg{Fn}{Function}{}{end}
	\Fn{GenerateCandidate($S_{k-1}$, Nbr)}{
		\KwData{Set of $(k-1)$-simplices $S_{k-1}$, Neighbor set $\text{Nbr}_v $ for each disk $v \in D$}
		\KwResult{Set of candidates for $k$-simplex}
		$C \gets \emptyset$;
		\For{$s$ \textbf{in} $S_{k-1}$} {
			$\hat{N} \gets \bigcap_{v \in s} \text{Nbr}_v$\;
			\For{$d$ \textbf{in} $\hat{N}$} {
				$C  \gets C \cup \{(s, d)\}$\;
			}
		}
		\Return{$C$}\;
	}
	\caption{Candidate Enumeration Algorithm}
	\label{algo:enum}
\end{algorithm}

\begin{algorithm}[htbp]
	\SetAlgoLined
	\SetKwProg{Fn}{Function}{}{end}
	\Fn{Verify($s_{k-1}$, $d$, $p$)}{
		\KwData{$(k-1)$-simplex $s_{k-1}$, disk $d$ that is shared neighbor of the vertices in $s_{k-1}$, Cached intersection $p$ of $s_{k-1}$}
		\KwResult{Intersection Point of the $k$-simplex $(s_{k-1},d)$}
		\If{the smallest disk is inside all others}
		{
			\Return{Center of the smallest disk}
		}
		\If{$p$ is inside $d$} {
			\Return{$p$}
		}
		\For{$d_i \in s_{k-1}$} {
			$X \gets \partial d_i\cap\partial d$\;
			\If{Any $pt \in X$ is inside all other disks}{
				\Return{$pt$}
			}
		}
		\Return{$\emptyset$}
	}
	\caption{Candidate Verification Algorithm}
	\label{algo:verify}
\end{algorithm}

\begin{algorithm}[htbp]
	\SetAlgoLined
	\SetKwProg{Fn}{Function}{}{end}
	\Fn{KSimplex($S_{k-1}$, $\text{Nbr}$)}{
		\KwData{Set of all $(k-1)$-simplices $S_{k-1}$, Neighbor set $\text{Nbr}_v $ for each vertex $v \in N$}
		\KwResult{Set of all $k$-simplices}
		$S_{k} \gets \emptyset$ \;
		$CachedP \gets \emptyset$ \;
		\For{$s_{k-1}, d \gets$ GenerateCandidate($S_{k-1}$, Nbr)}{
			$p \gets$ Verify($s_{k-1}$, $d$, CachedP[$s_{k-1}$])\;
			\If{$p$}{
				$s_k \gets s_{k-1} \cup \{d\}$ \;
				$S_k \gets S_k \cup \{s_k\}$ \;
				CachedP[$s_k$] $\gets p$ \;
			}
		}
	    \Return{$S_{k}$}\;
	}
	\caption{Construction of all $k$-simplices}
	\label{algo:simk}
\end{algorithm}

We adapt the ideas from above and construct the algorithm in two steps, which are candidate enumeration and candidate verification, as shown in Algorithm~\ref{algo:enum} and \ref{algo:verify}. Then we can use those two functions to compute $k$-simplices as shown in Algorithm~\ref{algo:simk}.

\subsection{3D Extension}
\label{subsec:algo3d}
To extend the algorithm to $\mathbb{R}^3$, we are given a set of balls $B$ in $\mathbb{R}^3$, where each ball $b_i \in B$ has its center located at $(x_i, y_i, z_i)$, with radius $r_i$. 
The candidate enumeration algorithm remains the same as of in the 2D case, while the candidate verification algorithm needs to be extended. 
The intersection for a $k$-simplex in $\mathbb{R}^3$ will be one of the following scenarios:
\begin{enumerate}
	\item The smallest ball which is inside all other $k$ balls,
	\item A single point,
	\item A region bounded by the surfaces of multiple balls.
\end{enumerate}

\begin{figure}[htbp]
	\centering
	\includegraphics[width=0.45\textwidth]{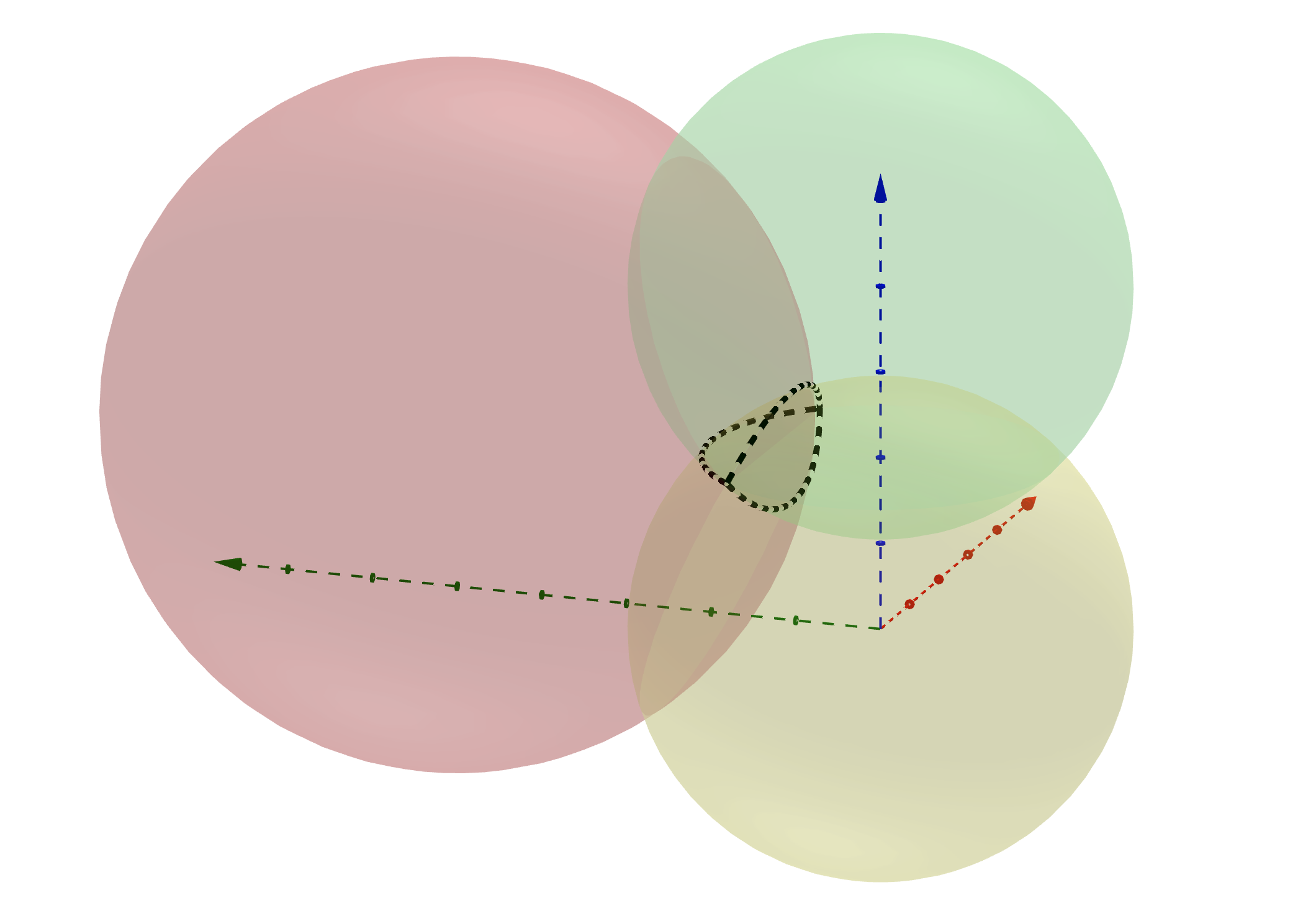}
	\caption{Intersection of three balls}
	\label{fig:crease}
\end{figure}

For these three scenarios, we will use a similar approach as what we did for the 2D setting. 
We can deal with the first scenario by checking if the center of the smallest ball is inside all other balls. Then for the other two cases, when a new ball is added to a $(k-1)$-simplex, we need to verify if they form a $k$-simplex, which will also lead to two cases: either the intersection region is completely contained in the new ball, or the new ball reduces the intersection region and thus belongs to the support set. 

When two balls intersect with each other, we get a union of two spherical caps. The spheres of two balls intersect with each other at a circle. This circle serves the same purpose as the intersection points for disks in 2D case, i.e. part of this circle is the ``crease'' of the intersection region. As shown in Figure~\ref{fig:crease}, three balls intersect, where the black dashed lines indicate the ``crease'' of the intersection region. This is the special part of the intersection region that is easier to compute.

If the intersection region is not affected, we can use a cached intersection point of the $k-1$-simplex to verify if it is the case. Otherwise, there has to be a ``crease'', which is formed by the newly added ball and some other balls and is contained by all balls in the $k$-simplex. Therefore, for each circle that is the intersection of the spheres of the newly added ball and another ball, we will intersect all other balls with the plane where this circle is, and verify that if there exists a non-empty intersection region between these intersected circles, similar to what we did in the 2D scenario. Eventually, we will get a point of the intersection region of the $k$-simplex that can be cached for the computation of ($k+1$)-simplex.

\begin{algorithm}[htbp]
	\SetAlgoLined
	\SetKwProg{Fn}{Function}{}{end}
	\Fn{Verify3D($s_{k-1}$, $b$, $p$)}{
		\KwData{$(k-1)$-simplex $s_{k-1}$, ball $b$ has non-empty intersection with every ball in $s_{k-1}$, Cached intersection $p$ of $s_{k-1}$}
		\KwResult{Intersection Point of the $k$-simplex $(s_{k-1},b)$}
		\If{the smallest ball is inside all others}
		{
			\Return{Center of smallest ball}
		}
		\If{$p$ intersects with $d$} {
			\Return{$p$}
		}
		\For{$b_i \in s_{k-1}$} {
			$c \gets \partial b_i\cap\partial b$ \;
			$f \gets$ plane where $c$ belongs to \;
			$C \gets \{c_j | c_j = \partial b_j \cap f\}$ \;
			\If{the smallest circle in $C$ is inside all others} {
				\Return{Center of the smallest disk}
			}
			\For{$c_i \in C$}{
			    if(any $p \in \{\partial c_i \cap \partial c\}$ is inside all other $c_j \in C$) {
			        return $p$ \;
			    }
			}
		}
		\Return{$\emptyset$}
	}
	\caption{Candidate Verification Algorithm in $\mathbb{R}^3$}
	\label{algo:verify3d}
\end{algorithm}

\subsection{Theoretical analysis}
\label{subsec:complexity}
First, we examine the time complexity of our algorithm, which is hard to analyze since the running time of this algorithm is output-sensitive. Our algorithm computes each dimension $k$ based on the result of the previous dimension $k-1$. We decompose the algorithm into two phases. 
In the candidate enumeration phase, each candidate we generate is essentially a Vietoris-Rips simplex that needs to be verified by finding a non-empty intersection among all disks. In the ideal case, every candidate we generate is indeed a \Cech simplex. Therefore, the number of candidates generated is lower bounded by the number of \Cech simplices and upper bounded by the number of Vietoris-Rips simplices. 
The running time of the candidate verification algorithm depends on the scenario of the intersection region for simplex $k$. If the new disk doesn't affect the intersection region, then it takes O(1) to check the cached intersection point. If the new disk is the smallest, then it takes O(k) to verify if its center is inside all other disks. The last scenario is when we need to compute the intersection points of the boundary of the new disk with all other disks in $k-1$-simplex (which is $O(k)$) and then verify if any of those intersection points is inside all other disks (which is $O(k)$ under 2D setting, and $O(k^2)$ under 3D setting). Therefore, the overall time complexity for verification is $O(1)$ in the best scenario, $O(k^2)$ in the worst case under the 2D setting, and $O(k^3)$ under the 3D setting.
To combine the two phases together, the time complexity to compute \Cech complex for dimension $k$ is $O(c_k)$ for the best scenario and $O(k^2\cdot vr_k)$($O(k^3\cdot vr_k)$ for 3D) for the worst scenario, where $c_k$ denotes the number of \Cech simplices in dimension $k$ and $vr_k$ denotes the number of Vietoris-Rips simplices in dimension $k$.

Secondly, we analyze the memory complexity. When we construct simplices for dimension $k$, we cached the intersection point for each simplex in dimension $k-1$ and $k$. Therefore, the overall memory complexity is $O(\max(c_k))$, where $\max(c_k)$ denotes the maximum number of simplices for one dimension $k$.

\subsection{Optimization for high dimensions}
\label{subsec:helly}
Helly's theorem\cite{bollobas2006art}(Problem 29) states that for a set of convex objects in $R^d$, if any $d+1$ members of the set have non-empty intersections, then the intersection of all members is also non-empty. Therefore, in 2D and 3D spaces, we only need to check up to 2-simplices and 3-simplices respectively, and the result in higher dimension is implied in the lower dimension. 

A direct application of Helly's theorem is that we can revise the verification algorithm for a candidate set $s_k$ for $k$-simplices where $k > d$ in $R^d$, if every subset of size $d+1$ forms a $d$-simplex, which can be easily verified from previous computation, then $s_k$ forms a $k$-simplex. However, this requires querying $\binom{k+1}{d+1}$ $d$-simplices, which is $O(k^3)$ in 2D space and $O(k^4)$ in 3D space. An extension of Helly's theorem is that for a set of $k+1$ convex objects $s_k$ where $k > d$, if any subset of $k$ convex objects has non-empty intersection, then any subset of $d+1$ objects has non-empty intersection, and therefore all members of the set $s_k$ have non-empty intersection. To verify a candidate set $s_k$ for $k$-simplices where $k > d$ in $R^d$, we could check if all of it's $k-1$ face are all in the complex. This will only require $O(k)$ of set queries. We will demonstrate the performance of those two approaches in Section~\ref{sec:exp}.

\section{Experiment and Performance Evaluation}
\label{sec:exp}
In this section, we are going to evaluate the performance of our algorithm under 2D setting in comparison to the work in~\cite{le2015construction}. We also compare our algorithm with the minimum enclosing ball algorithm under both 2D and 3D settings but for standard \Cech complex.
All algorithms are implemented in Python 3.8 and running on a machine with a 2GHz Intel i5 CPU and 16GB memory. 

\begin{table*}[!ht]
	\begin{tabular}{ll|ll}
	\multicolumn{2}{c}{Datasets}                            & \multicolumn{2}{c}{Running Time} \\
	Set Size & Density and distribution   & Our method        & Le et al. \\
	\hline
	40   & Evenly                     & 1.993 ms ± 0.080 ms & 40.505 ms ± 1.123 ms\\
	40   & Random                     & 10.863 ms ± 0.458 ms  & 865.707 ms ± 13.439 ms\\
	90   & Evenly                     & 31.506 ms ± 3.207 ms  & 14.931 s ± 696.922 ms \\
	90   & Random                     & 153.040 ms ± 2.115 ms & 3.947 min ± 3.908 s\\
	150  & Evenly & 69.367 ms ± 5.594 ms & 1.439 min ± 1.291 s\\
	150  & Random & 554.581 ms ± 19.980 ms  & 74.689 min ± 2.222 min\\
	10000 & Evenly & 1min 27s ± 1.4 s & 175.63 min ± 2.89 min \\
	& & &(stops early at $k=6$)
	\end{tabular}
	\caption{Running time Comparison for constructing 2D generalized \Cech complex}
	\label{table:runtime}
\end{table*}
	
\begin{figure*}[htbp]
	\centering
	\begin{subfigure}[b]{0.3\textwidth}
		\centering
		\includegraphics[width=\textwidth]{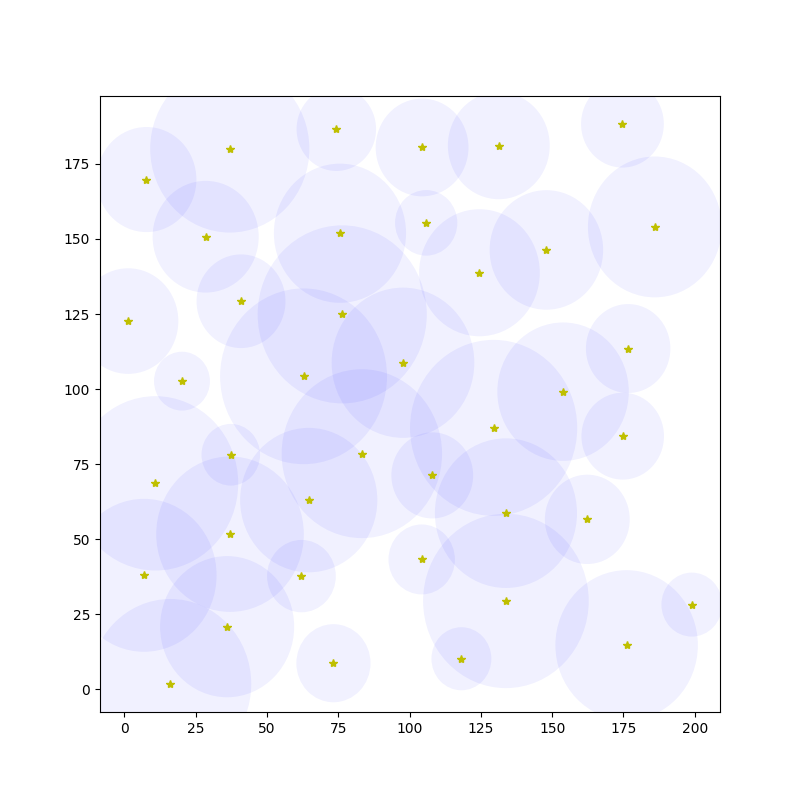}
		\caption{40 Disks Evenly distributed}
		\label{fig:40even}
	\end{subfigure}
	\hfill
	\begin{subfigure}[b]{0.3\textwidth}
		\centering
		\includegraphics[width=\textwidth]{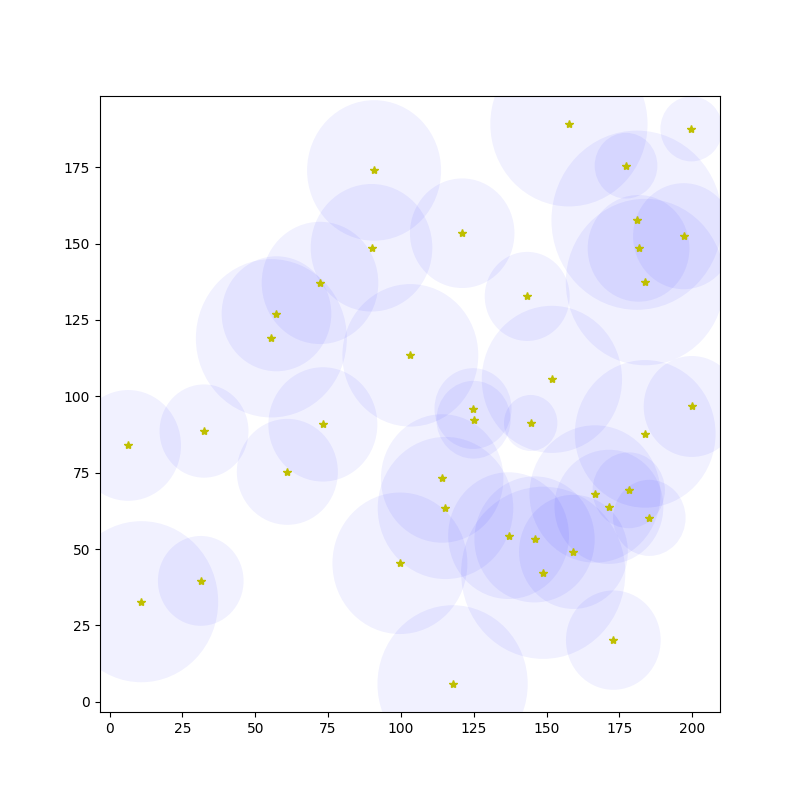}
		\caption{40 Disks Randomly distributed}
		\label{fig:40uneven}
	\end{subfigure}
	\hfill
	\begin{subfigure}[b]{0.3\textwidth}
		\centering
		\includegraphics[width=\textwidth]{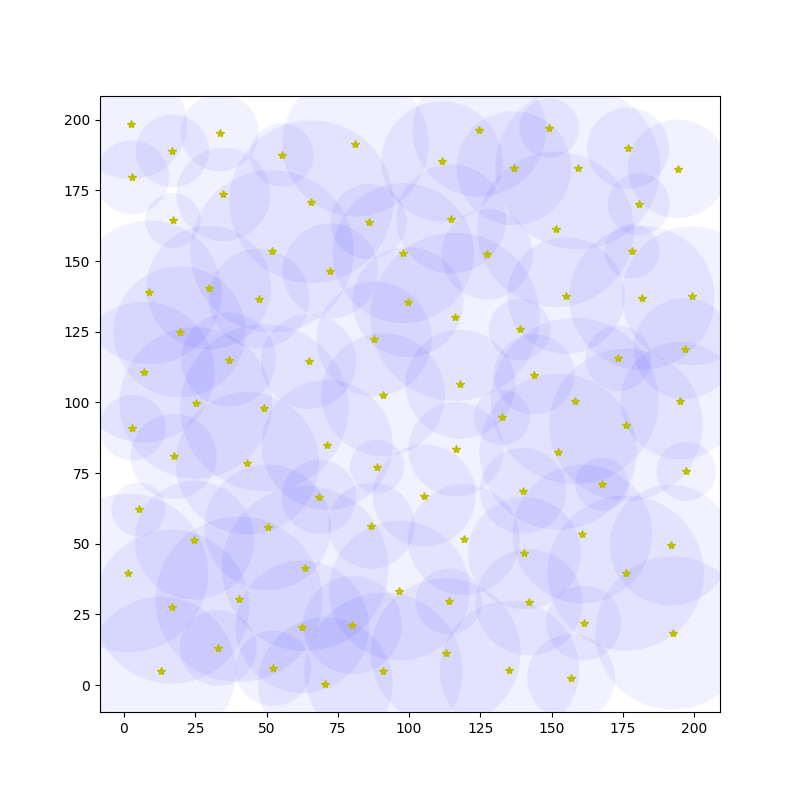}
		\caption{90 Disks Evenly distributed}
		\label{fig:90even}
	\end{subfigure}
	\hfill
	\begin{subfigure}[b]{0.3\textwidth}
		\centering
		\includegraphics[width=\textwidth]{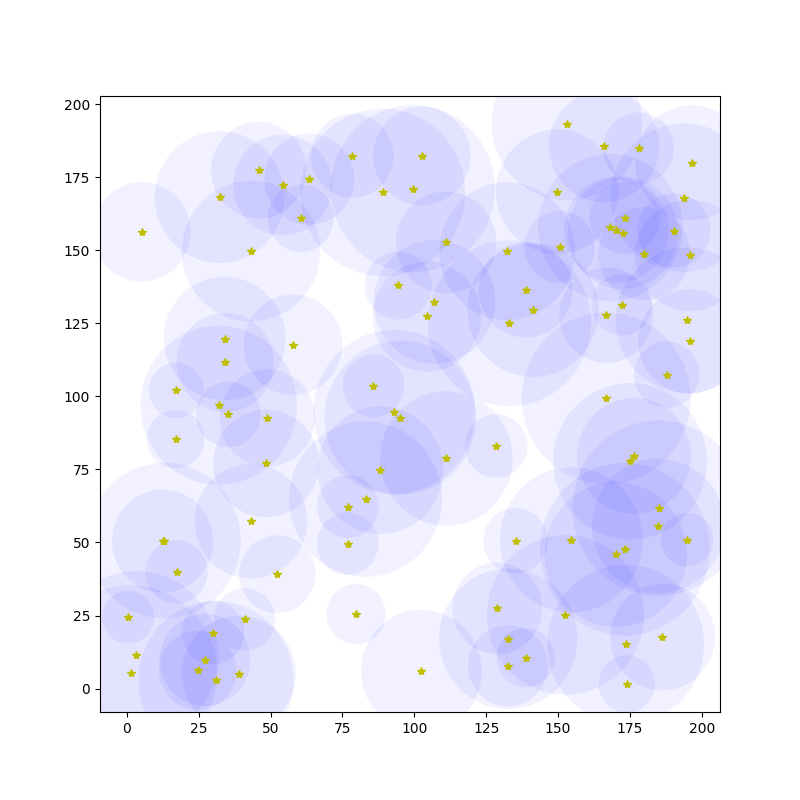}
		\caption{90 Disks Randomly distributed}
		\label{fig:90uneven}
	\end{subfigure}
	\hfill
	\begin{subfigure}[b]{0.3\textwidth}
		\centering
		\includegraphics[width=\textwidth]{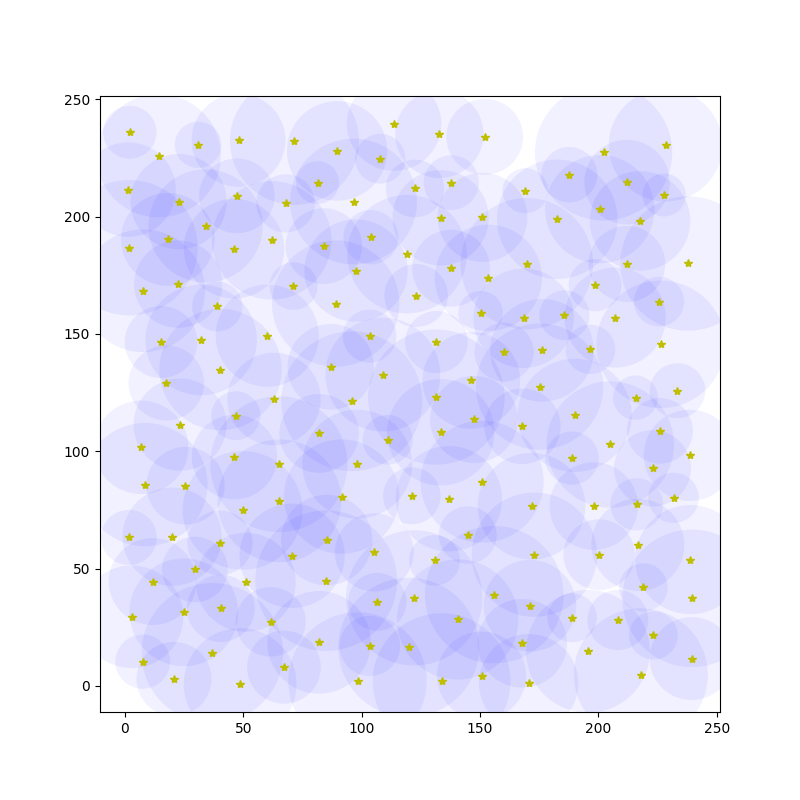}
		\caption{150 Disks Evenly distributed}
		\label{fig:150even}
	\end{subfigure}
	\hfill
	\begin{subfigure}[b]{0.3\textwidth}
		\centering
		\includegraphics[width=\textwidth]{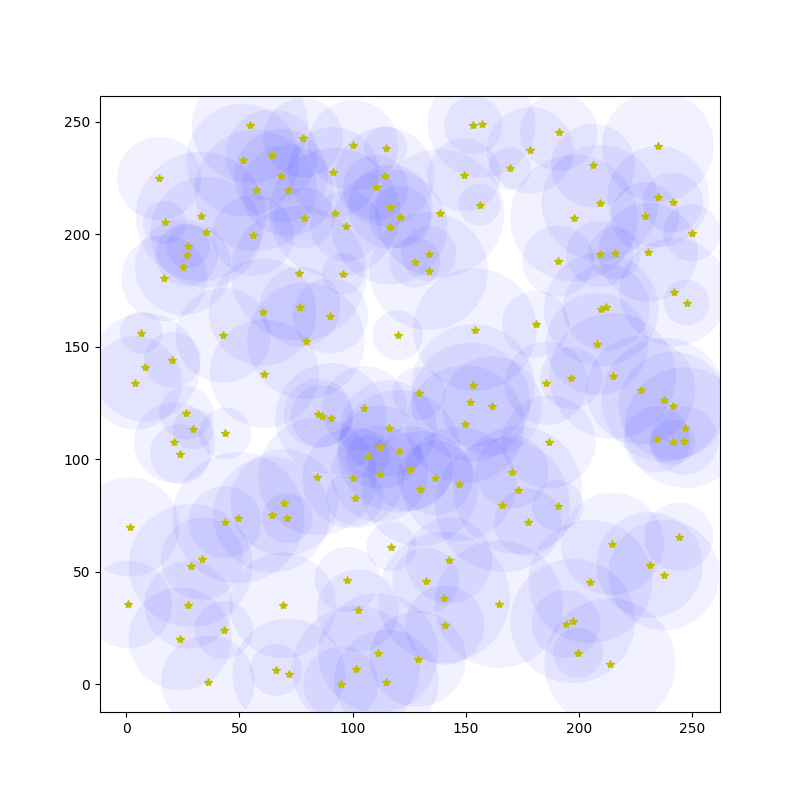}
		\caption{150 Disks Randomly distributed}
		\label{fig:150uneven}
	\end{subfigure}
	\hfill
	\caption{Experiment Dataset}
	\label{fig:dataset}
\end{figure*}
	
\begin{figure}[htbp]
	\centering
	\includegraphics[width=0.45\textwidth]{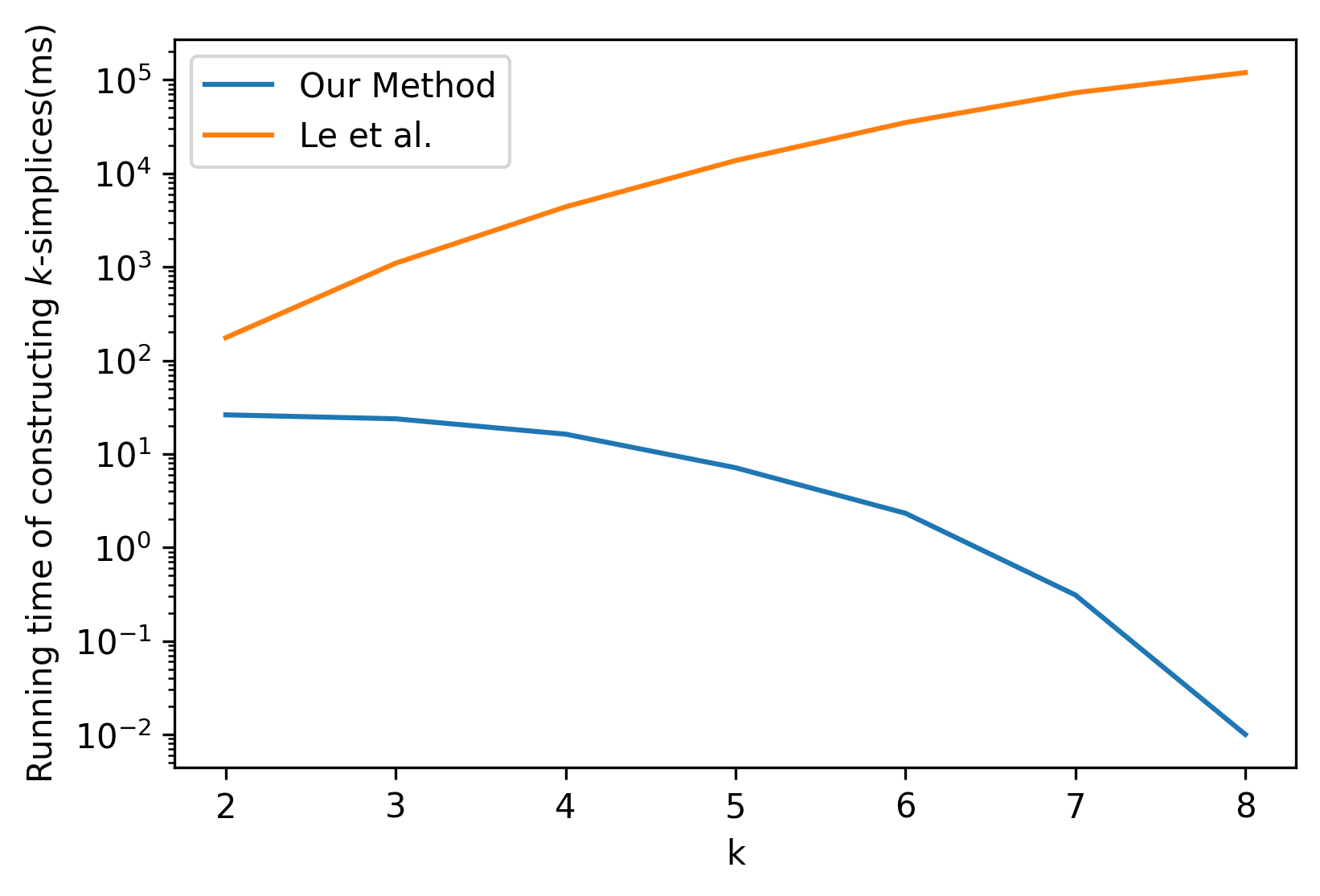}
	\caption{Comparison of running time for each dimension for data shown in Figure~\ref{fig:90uneven}}
	\label{fig:runtimek}
\end{figure}

\begin{figure}[htbp]
	\centering
	\includegraphics[width=0.45\textwidth]{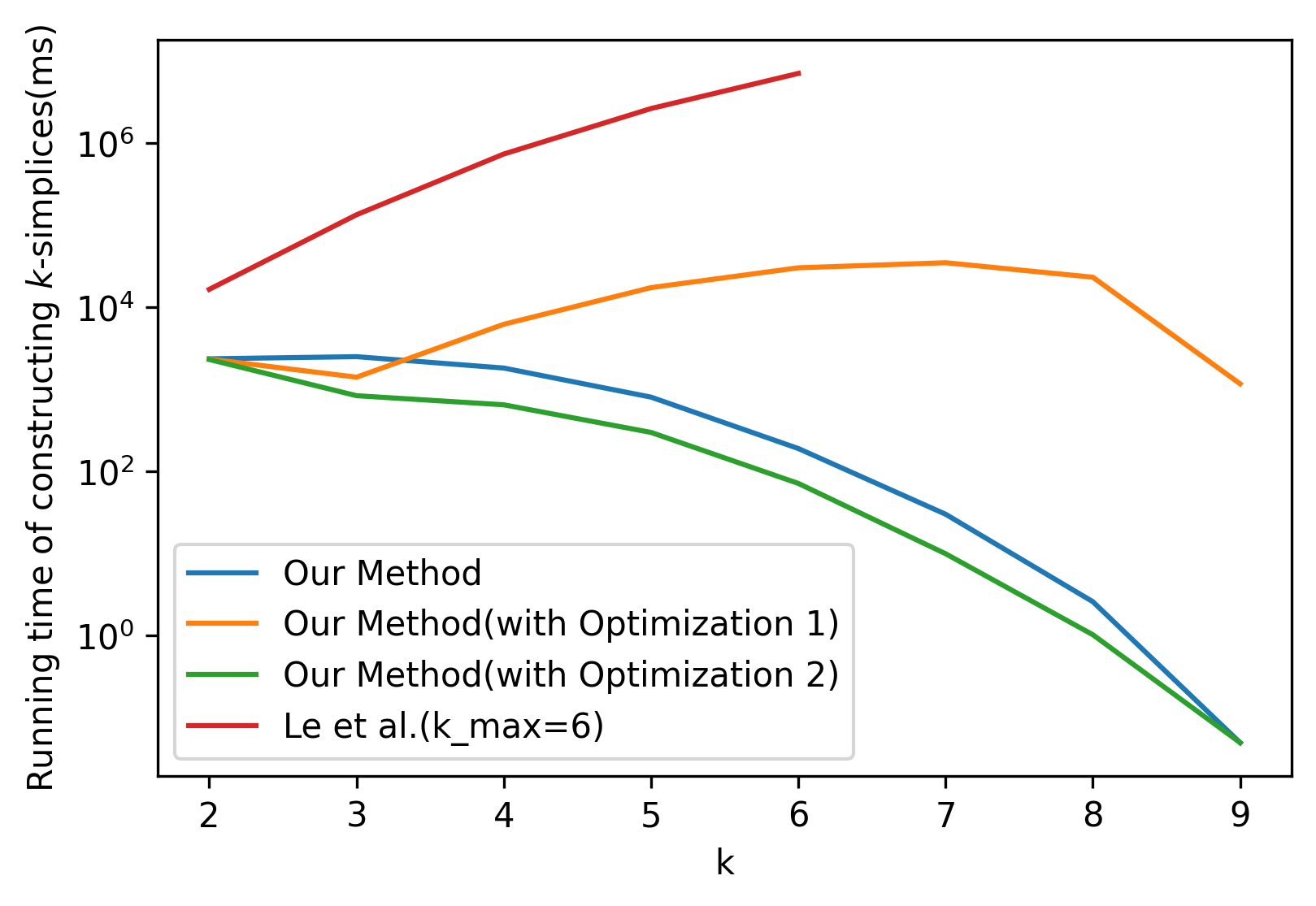}
	\caption{Comparison of running time for each dimension for dataset with 10k disks}
	\label{fig:optim}
\end{figure}

\begin{figure}[htbp]
	\centering
		\begin{subfigure}[b]{0.45\textwidth}
			\centering
			\includegraphics[width=\textwidth]{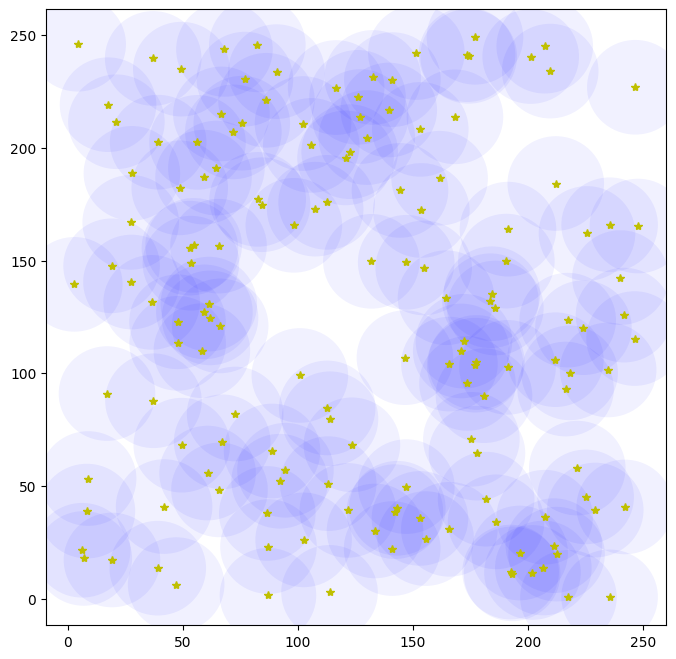}
			\caption{150 Disks Randomly distributed with same radius}
			\label{fig:150standard}
		\end{subfigure}
		\hfill
		\begin{subfigure}[b]{0.45\textwidth}
			\centering
			\includegraphics[width=\textwidth]{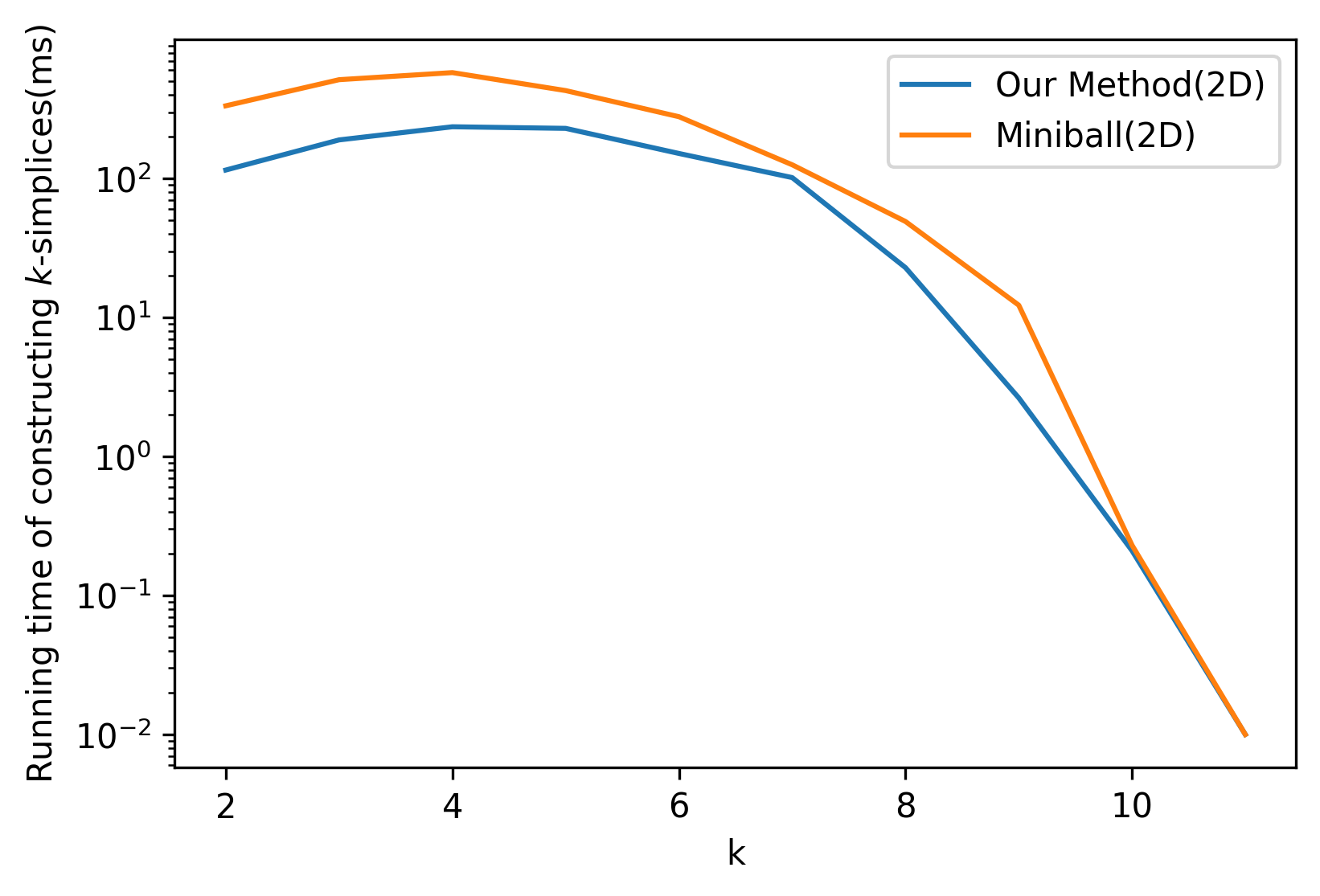}
			\caption{Comparison of running time for each dimension}
			\label{fig:runtimeminiball}
		\end{subfigure}
		\caption{Comparison of running time with Minimum Enclosing Ball algorithm for 150 disks with same radius}
	\label{fig:runtimemb}
\end{figure}

\begin{figure}[htbp]
	\centering
	\begin{subfigure}[b]{0.45\textwidth}
		\centering
		\includegraphics[width=\textwidth]{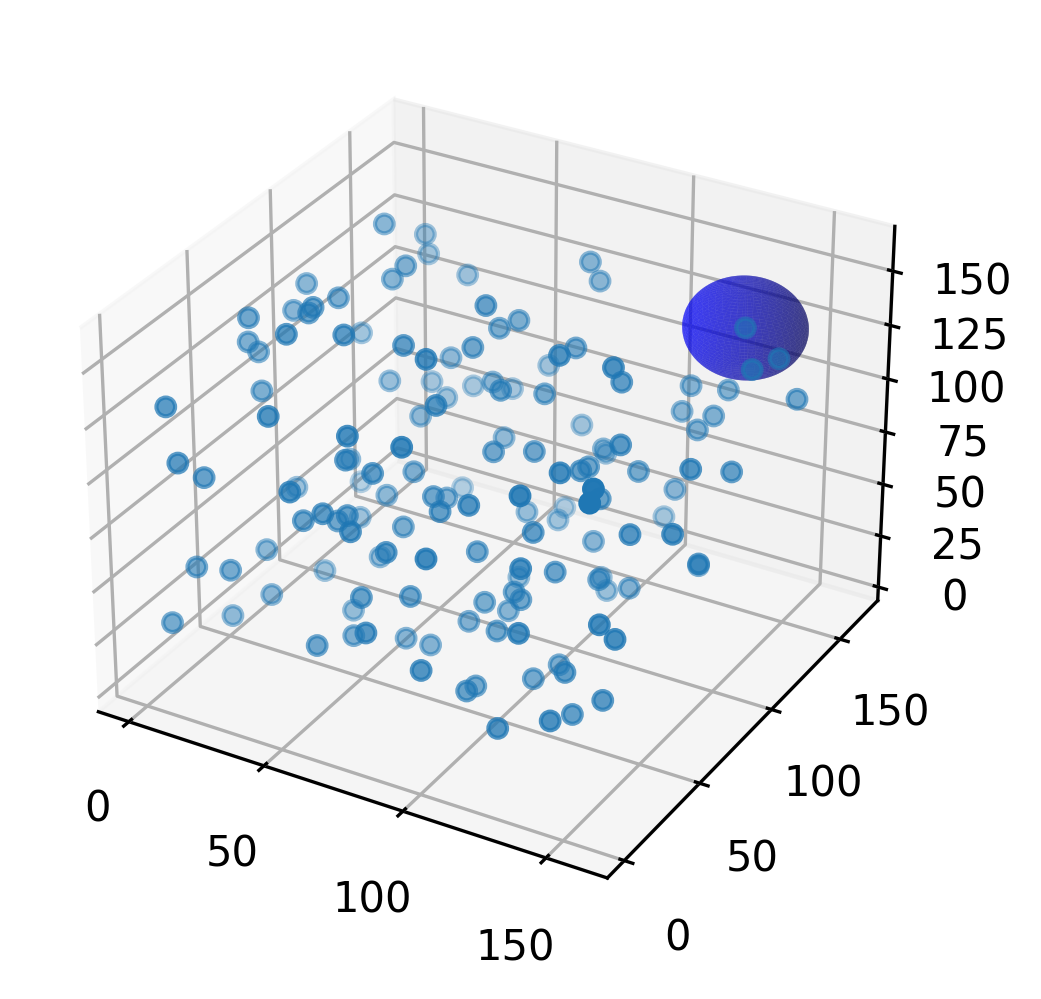}
		\caption{150 Balls Randomly distributed with same radius}
		\label{fig:150standard3d}
	\end{subfigure}
	\hfill
	\begin{subfigure}[b]{0.45\textwidth}
		\centering
		\includegraphics[width=\textwidth]{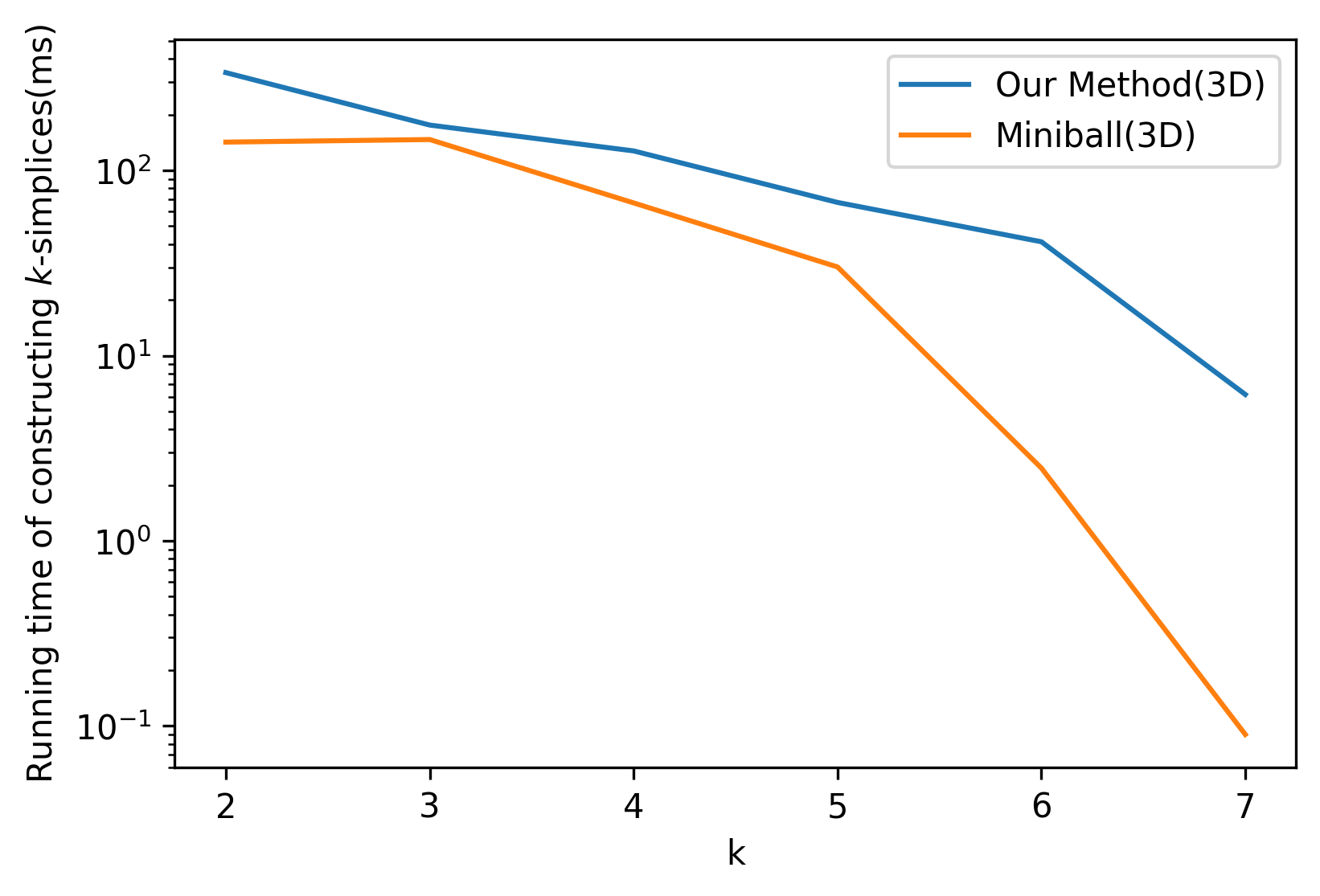}
		\caption{Comparison of running time for each dimension}
		\label{fig:runtimeminiball3d}
	\end{subfigure}
	\caption{Comparison of running time with Minimum Enclosing Ball algorithm for 150 balls with same radius under 3D setting}
	\label{fig:runtimemb3d}
\end{figure}

As shown in Figure~\ref{fig:dataset}, we tested the two algorithms with both evenly distributed and randomly distributed sets of disks, that have various sizes and densities. We also generate a dataset with 10k disks evenly distributed, which follows a similar density as seen in Figure~\ref{fig:150even} but is not plotted due to the size of the dataset. We use the Poisson Disk Sampling method described in \cite{bridson2007fast} to generate positions for the evenly distributed disks and use uniform distribution to generate the randomly distributed disks. The radius of each disk is generated as $\alpha\cdot R$, where $R$ is a fixed number and $\alpha$ is a scale factor that follows a uniform distribution $U(0.3, 1)$. The running time is an average of 7 runs on the same data set, as shown in Table~\ref{table:runtime}. 

We observe here that our algorithm outperforms in every setting,
achieving up to thousands of times boost in performance. The performance margin grows as the density and size of the data set to grow. The improvement of performance comes from two aspects. The first one, when we generate the candidates for potential $k$-simplices, only the $(k-1)$-simplices are considered instead of all combinations of $k$ neighbors of one disk. Fewer $(k-1)$-simplices leads to fewer $k$-simplex candidates for verification. The second part comes from the candidate verification algorithm, where we don't need to enumerate all pairs among $k+1$ disks to verify if any of their intersections points lies in all others, and thus we reduce the running time from $O(k^3)$ to $O(k^2)$ for the worst scenario when verifying a single $k$-simplex candidate. Therefore, if the disks are densely located which leads to a higher dimension in the \Cech complex, that is where our algorithm outperforms by a large margin. 

For example, as shown in Figure~\ref{fig:90uneven}, the disks are randomly deployed with areas of high density and thus result in high dimensions in \Cech complex. Our algorithm can complete the computation in just 153 ms, in comparison, the algorithm in~\cite{le2015construction} will take around 4 minutes to compute all 8 dimensions. Figure~\ref{fig:runtimek} gives a better view of the details of the running time. Here, we compare the running time to construct all $k$-simplices for each $k$, where $k \in [2, 8]$. While the running time of the algorithm in~\cite{le2015construction} grows sub-exponentially with the dimension $k$, the running time of our algorithm remains on a very low level. The running time of our algorithm drops as the number of simplices in higher dimensions reduces.

We also evaluate the efficiency of the optimization for simplices with higher dimensions, which we proposed in Section~\ref{subsec:helly}. As shown in Figure~\ref{fig:optim}, the performances of three versions of our algorithm are evaluated with the 10k dataset. Optimization 1 denotes the algorithm with direct application of Helly's theorem and Optimization 2 denotes the extended version. As we expected, the direct application of Helly's theorem doesn't perform well due to its cubic time complexity, whereas the extended version does improve the performance by 60-80\% at certain dimensions. We also give the result from \cite{le2015construction} for reference, which stops early at $k=6$ and yet is still outperformed by two orders of magnitude.

In order to compare our algorithm with \cite{dantchev2012efficient}, we generated 150 disks(/balls) with equal radius as shown in Figure~\ref{fig:150standard} and Figure~\ref{fig:150standard3d}.
For the 3D dataset, we only show the positions of the center of each ball for clarity, and one sphere is plotted to demonstrate the size of each ball. We replaced the verification algorithm with the technique used in \cite{dantchev2012efficient}, that the minimum enclosing ball of the centers of $k+1$ disks is calculated, and they formed a $k$-simplex only if the radius of the minimum enclosing ball is less than the radius of all disk. We use the implementation of the Minimum Enclosing Ball algorithm provided in the Python library cechmate~\cite{cechmate}. As shown in Figure~\ref{fig:runtimeminiball}, we achieve better performance in the process of verifying the intersection of $k$-simplex and under the 2D setting. While under 3D setting, as shown in \ref{fig:runtimeminiball3d}, our algorithm is slightly slower, primarily due to the time complexity of candidate verification algorithm increases from $O(k^2)$ to $O(k^3)$ and additional processing such as projecting all balls to a 2D plane. However, our algorithm is capable of constructing the standard \Cech complex when the radius of each circle(/ball) is different. 

Both Gudhi\cite{gudhi-software} and Cechmate\cite{cechmate} library provides functions related to \Cech complex construction. Gudhi doesn't expose the \Cech complex class in their python interface, and it can only compute standard \Cech complex.
Cechmate is implemented in python, but it only computes standard \Cech complex filtration which makes it hard to compare the performance directly, plus it is also based on the minimum enclosing ball algorithm. Therefore, we did not evaluate the performance of our algorithm against those two libraries, while we do compare against the minimum enclosing ball algorithm which both libraries rely on.

\section{Conclusion}
\label{sec:conclusion}
In this paper, we present an algorithm to construct generalized \Cech complex that works on set of 2D disks with various radius. This algorithm is improved over the work in \cite{le2015construction}, and we evaluate the performance of both algorithms which shows that our algorithm performs much better in every scenario, especially when dealing with a large data set that results in higher dimension in the \Cech complex. We also extend the algorithm for 3D settings, expanding the domain of potential applications. For future work, it is possible to design a parallel algorithm that utilizes the power of GPU to accelerate construction. A different approach to constructing generalized \Cech complex is via weighted minimum enclosing ball algorithm since they are closely related. We will also explore the application of the \Cech complex for modeling the mobile sensor network where the \Cech complex retains the geometric information the best.  

\section*{References}

\printbibliography

\end{document}